\documentstyle[12pt,epsf,epsfig,wrapfig]{article}

\begin{document}

\begin{center}

{\large \bf Neutron spin-dependent structure function,}

{\large \bf Bjorken sum rule, and first evidence}

{\large \bf for singlet contribution at low $x$  }

\vspace{5mm} 

\underline{J. Soffer}$^1$,  and O. Teryaev$^2$

\vspace{5mm} 

{\small\it (1) Centre de Physique Th\'eorique - CNRS - Luminy,\\
Case 907 F-13288 Marseille Cedex 9 - France\\
(2) Bogoliubov Laboratory of Theoretical Physics, \\
Joint Institute for Nuclear Research, Dubna, 141980, Russia \\
} 
\end{center}

\begin{center} ABSTRACT  

\vspace{5mm}

\begin{minipage}{130
mm} \small We perform the isospin decomposition of proton and neutron SLAC data
 in the region $0.01 \leq x \leq 0.1$.
The isovector part is well described by a power behaviour
$x^{\alpha}$, where $\alpha$  leads to the validity of Bjorken sum rule and it
is consistent with the power extracted from all previous data using NLO
evolution. The isoscalar part behaviour may be interpreted as
a partial cancellation between a positive non-singlet contribution
and a singlet one strongly negative.
\end{minipage} 

\end{center}

The very accurate measurement of the neutron spin-dependent
structure function $g_1^n$[1], whose results have been presented
recently, possess two remarkable
properties. First, the values of neutron structure function at $Q^2=5GeV^2$
are rather large
and negative in the region of moderately low $x$. Second, the data can be
rather accurately fitted by the power function $x^{-0.8}$ and this power
seems to be unexpectedly large. It is not obvious, where this
large number, for which there is no
indication in the proton data, is coming from,
so it may affect the extrapolation to $x=0$ and cast some doubts on
the validity of Bjorken sum rule.
In the present work we perform the isospin decomposition of the data
for both proton and neutron.
As a result, we conclude that the isovector contribution is well approximated
by the power behaviour found earlier by an elaborate method based on
next-to-leading order (NLO)
evolution[2]. It may be also interpreted  as the manifestation
of $ln^2x$ terms[3]. For the
isoscalar part one has the
signature for a rather singular singlet contribution, compatible
with the similar behaviour predicted by QCD[4].
This may imply a significant gluon polarization in the nucleon in this
range of $x$.

The main feature of the new neutron data[1]
is large and negative $g_1^n \sim -g_1^p$,
measured with a good accuracy
up to small $x$, say $x \sim 0.01$.
The SLAC proton data[5] at $Q^2=3 GeV^2$ are positive and of roughly
the same magnitude, but on the
contrary, rather flat in this region.
This could be understood, qualitatively, as a result of the
interplay of a negative contribution at low $x$, responsible for
the singular behaviour of the neutron, and a
positive contribution at larger $x$.
To check this assumption it is instructive to consider the isovector
contribution. Since there is no clear evidence of scaling violations
between $Q^2=3 GeV^2$ and $Q^2=5 GeV^2$ in any
polarized deep inelastic scattering experiment, we may neglect, for the time
being, the effects of QCD evolution,
because we are only interested in the results, provided by the data at the
present level of statistical accuracy.
It is then possible, by combining the SLAC neutron data with the
SLAC proton data, to determine the quantity $g_1^{p-n} \equiv g_1^p-g_1^n$,
entering the Bjorken sum rule, with a higher accuracy, than for proton alone.
This is because the neutron data are fortunately negative, so the
difference is larger in magnitude than $g_1^p$, while the errors are
 practically the same for $g_1^{p-n}$ and $g_1^p$, given
the quality of the neutron data.

Performing this analysis of the data, we see immediately a behaviour less
sharp than that of neutron.
Since phase space effects, as powers
of $(1-x)$, are not so important in the region under consideration, we
were looking for a simple power parametrization of the data,
implied by Regge pole behaviour,
and we found that
this works rather well, for $0.016 \leq x \leq 0.125$ with
 the following expression

\begin{equation}
g_1^{p-n}(x)=0.147 x^{-0.45},
\end{equation}
as shown in Fig.1.

\begin{wrapfigure}{r}{7cm} \epsfig{figure=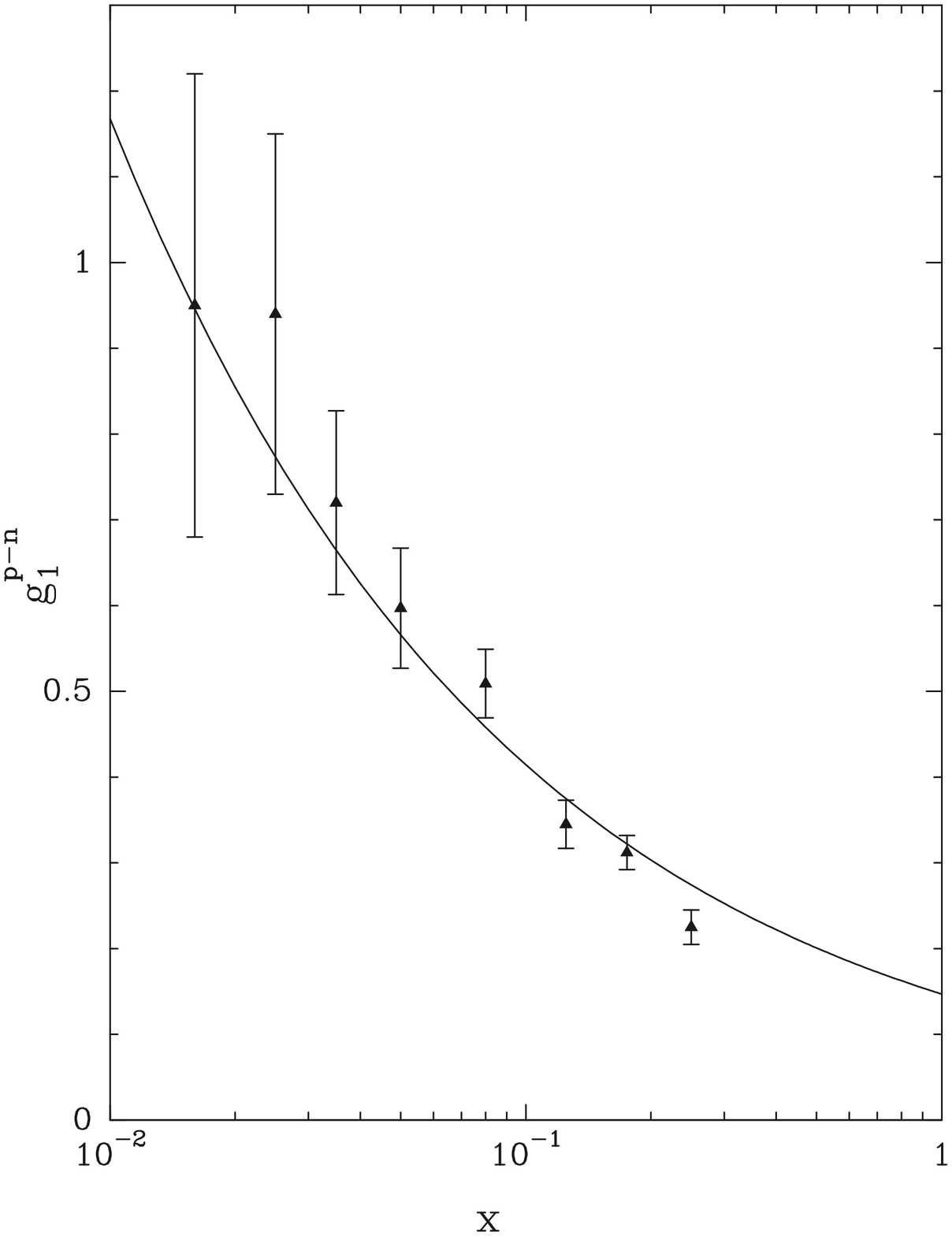,width=7cm} 
{\small
Figure 1: Comparison for $g_1^{p-n}$ between the curve given by eq.(1)
and the SLAC data refs.[1,5].} 
\end{wrapfigure}

The power we obtain is significantly smaller than
the expected contribution of the $a_1(1260)$ meson trajectory $(\sim 0.14)$.
As a result, the contribution to the Bjorken
integral from the region $0 \leq x \leq 0.125$ is large

\begin{equation}
\int_{0}^{0.125}dx g_1^{p-n}(x)=0.085.
\end{equation}

The region of higher
$x$ corresponds to a neutron contribution much smaller than that
of the proton, the latter also providing a large contribution to the above
integral, which is $0.09$ for $0.125<x<1$.
The total contribution to the Bjorken sum rule, obtained in this way, i.e.
$0.175$, 
appears to be in the fair agreement with the theoretical value.
Note that when the final neutron data will be available, as well as more
precise proton data, it will allow  a more serious analysis, taking also
into account the effects of QCD evolution.

\vspace{0.3cm}
At the present moment we would like to stress, that by combining
the current neutron and proton data, one is led to good agreement with
Bjorken sum rule.
Let us stress that this power is compatible with the one obtained using
NLO fit[2] to all previous data
$(-0.56 \pm 0.21)$. It is consistent with the result of $ln^2x$ summation[3].

%%%%%%%%%%%%%%%%%%

Since we found, that the sharp neutron structure function is not seen
in the difference between proton and neutron, it should be attributed
to the isoscalar channel. Also, the partial cancellation, we
suspect to be at the
origin of a flat proton structure function, should be manifested
in this channel as well. To check this, we calculated the quantity
$g_1^{p+n} \equiv g_1^p+g_1^n$. It really shows a
rather flat structure for $x \geq 0.035$ which should be related to the
interplay of the negative sharp contribution, showing itself in the fit
$x^{-0.8}$ to the neutron data[1],
and a positive contribution with a smaller power, dominating at larger $x$.
Since there is no counterpart for such a sharp behaviour in the conventional
Regge analysis, we make a strong, but natural assumption,
 that it
manifested itself in the $SU(3)$-singlet channel.
It is in this channel
that a strong mixing between polarized quarks and gluons provides the
anomalous gluon contribution to the first moment of $g_1$[6].

 At low $x$
the quark-gluon mixing
provides a strong correction to the subleading behaviour[4],
producing a power close to 1.
For a first estimate we  find that the data are
well described by 

\begin{equation}\label{p+n}
g_1^{p+n}(x)=0.145 x^{-0.45}-0.03 x^{-0.87},
\end{equation}
as seen in Fig.2.

\begin{wrapfigure}{r}{7cm} \epsfig{figure=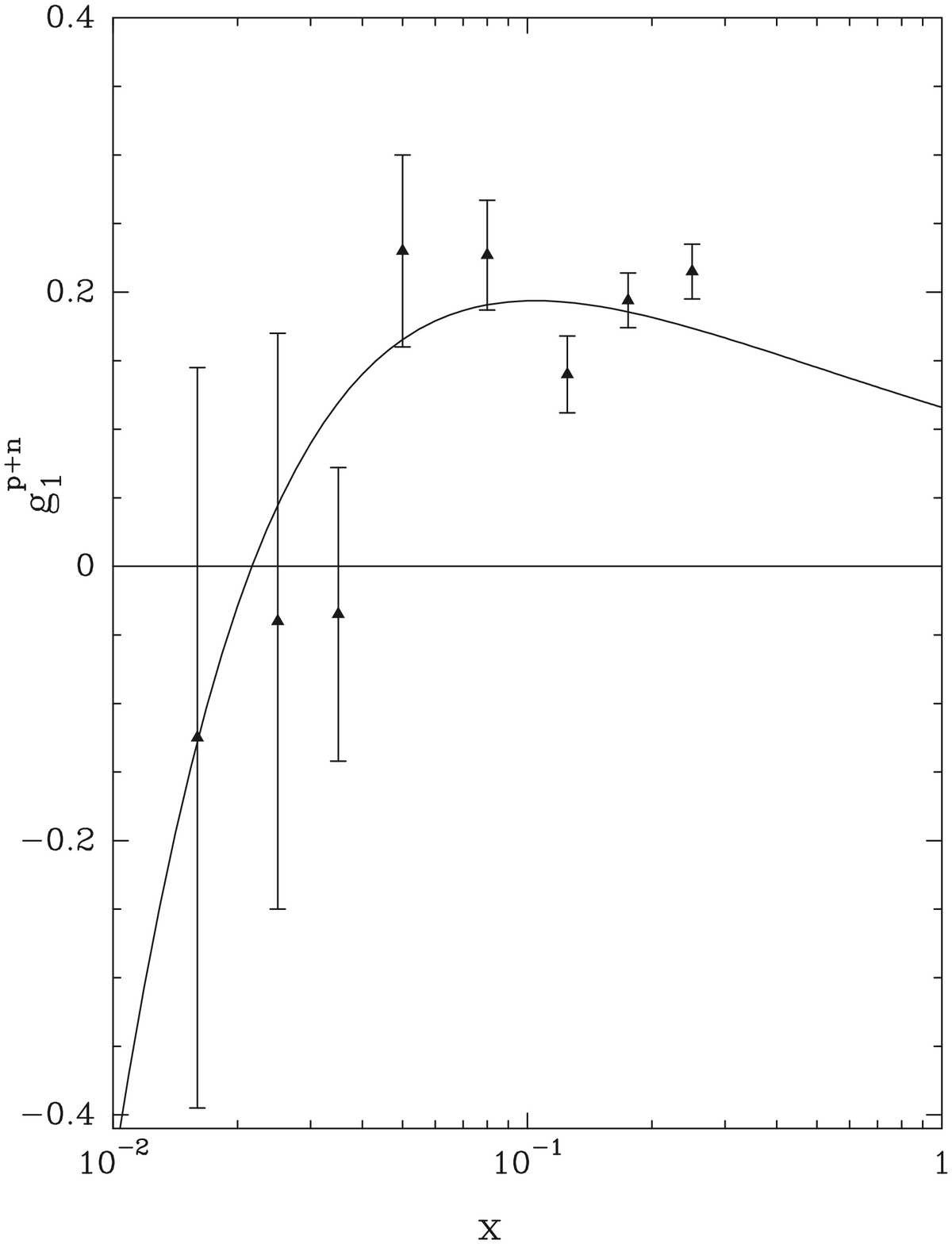,width=7cm} 
{\small
Figure 2: Same as Fig.1 for $g_1^{p+n}$
with the curve given by eq.(3).} 
\end{wrapfigure}

This formula is suggesting that the isoscalar contribution is approximately
equal to the isovector one, which is not so surprising, in order to have
the neutron structure function, dominated
by the most singular power only. Note that eq.(3) gives also a fair
description of the $g_1^d$ data in this kinematical region[7] and
further details on this result are given in ref.[8].

\vspace{0.2cm}

%\begin{wrapfigure}{r}{8cm} \epsfig{figure=fm.ps,width=8cm} 
%{\small
%Figure 1: Comparison for $g_1^{p-n}$ between the curve given by eq.(1)
%and the SLAC data refs.[1],[5].} 
%\end{wrapfigure}

%\begin{wrapfigure}{r}{8cm} \epsfig{figure=fp.ps,width=8cm} 
%{\small
%Figure 2: Same as Fig.1 for $g_1^{p+n}$
%with the curve given by eq.(7).} 
%\end{wrapfigure}

\vfill

\baselineskip 12pt

{\small\noindent[1]
E. Hughes, (SLAC experiment E154), presented at ICHEP Warsaw, July
1996.

\noindent[2] R.D. Ball, S. Forte and G. Ridolfi, Phys. Lett. {\bf B378}
(1996) 255; hep-ph/9608399.

\noindent[3] J. Bartels, B.I. Ermolaev and M.G. Ryskin, Z. Phys. {\bf C70}
(1996) 273.

\noindent[4]
J. Bartels, B.I. Ermolaev and M.G. Ryskin, hep-ph/9603204, March 1996.

\noindent[5] K. Abe et al. (SLAC experiment E143),
 Phys. Rev. Lett.{\bf 74} (1995) 346.

\noindent[6] A.V. Efremov, J. Soffer, O.V. Teryaev,
Nucl. Phys. {\bf B346} (1990) 97.

\noindent[7] K. Abe et al. (SLAC experiment E143),
 Phys. Rev. Lett.{\bf 75} (1995) 248.

\noindent[8] J. Soffer, O.V. Teryaev,
Preprint CPT-96/P.3387, hep-ph/9609329.
}

\end{document}